# Bayesian Optimized Event Based Epidemic Modeling in India

Olivi Thykkoottathil James*, Sinnu Susan Thomas

*Optimization and Machine Learning Lab, Indian Institute of Information Technology Management Kerala, Trivandrum India, 695581.*

**Abstract**

Pandemic outbreak creates a life threatening situation around the world and for a while it feels as if it slows down the world. A lot of effort is being taken to mitigate the spread of the pandemics. The main objective of this paper is to build a mathematical model of the pandemic spread in India based on the mass gathering using distribution functions and Bayesian approach. Subsequently modeled the pandemic spread based on the flights arrived in India using generalized linear regression. We validated the effect of these events in the number of confirmed cases in India and formulated the number of confirmed cases in the positive sense. Subsequently, studied the progression of the infective using progression series and flattened the curve using an appropriate convergence criterion. We validated the theoretical aspects with the statistics released by the Government.

## 1. Introduction

An epidemic named COVID-19 is an infectious disease caused by a new virus called Severe Acute Respiratory Syndrome Coronavirus 2 (SARS-CoV-2) that emerged in the city of Wuhan, China during December 2019 World Health Organization (2020a) and eventually the disease spread to other parts of the world. It was declared as a Public Health Emergency of International Concern on $30^{th}$ January, 2020 by WHO World Health Organization (2020b).

*Corresponding author

These are a large family of viruses that can cause illness, ranging widely in severity. Initially the coronavirus or COVID-19 was infected people through open-air and later it spread rapidly through direct contact with an affected person or an object and these carriers of infection flew from one part of the world to the other. More than 32 million people have been infected in the world and in that 994 thousand people have been killed Worldometer (2020). India is the front runner of this race with more than 5.9 million people infected cases and 94 thousand death cases. These viruses have different life span on different surfaces. COVID-19 tends to affect people with weak immune systems or the people already suffering from other diseases. This does not mean that people with a good immune system are invincible. Major symptoms of this pandemic are dry cough, fever, throat pain, tiredness, aches, runny nose, sore throat and difficulty in breathing Centers for disease control and prevention (2020). Since the symptoms appear in two-fourteen days, silent spreading of cases is at high.

Even though the world has seen many pandemics, it has never affected the entire world to this extent. Earlier pandemics were limited to some parts of the world. In 2009, a virus called novel influenza A (H1N1) was first re- ported in the United States. It spread rapidly in the United States and to other countries also. In June 2012 the first case of novel Middle East Respiratory Syndrome coronavirus (MERS-CoV) was reported in Jeddah, Saudi Arabia in Arabian Peninsula. About 35% of the infectives have died. The infection resulted from zoonotic transmission from bats and/or camel most likely World Health Organization (2019). Globally researchers are in search of medicines for the virus. This vast problem encouraged us to analyze the spread of pandemic under various situations. Various artificial factors are a cause of spread of this pandemic and in this paper we study various factors aggravating the spread of the pandemic. In India some of the events paved a way for the spread of the pandemic. Many mass gatherings paved a way for the spread of the pandemic across the country. Another big milestone was to bring back the stranded Indians from other countries, increasing the number of infectious cases.

While studying the effects of spread of the pandemic, the spread exhibited a geometric series. The conditions on termination of the series is formulated to study the flattening of the spread curve.

The contribution of this paper is three-fold given as:

- Developed an epidemic model based on mass gathering using Bayesian

inference and flight data using regression models.

- Proposed a model using geometric progression for flattening the curve of the pandemic.

This paper is organized into four sections. Section 2 explains different state-of-the-art approaches that have been carried out before and during the pandemic. Section 3 talks about the modeling of the pandemic for the mass gathering and flight data. Conditions on the spread leading to flattening is studied in this section. Section 4 discusses the results of the model developed. Lastly, the final section concludes the paper.

## 2. Literature Review

History records humans have gone through many pandemics in the world. Study suggests that some of the pandemic were caused due to the viruses from the zoonotic reservoirs like camels or bats Salata et al. (2020). A lot of work has been carried out regarding different pandemic and various models exist to study and take remedial actions to control the pandemic.

### *2.1. Modeling COVID-19*

COVID-19 outbreak was reported in Mainland China during December 2019. The study Zhong et al. (2020) has conducted experiments to predict the spread of COVID-19 using the Susceptible-Infected-Removed (SIR) model and it was conducted in the early stage of Covid-19 outbreak. The study concluded the linear regression model is inappropriate for fitting the infection rate of COVID-19. The study of COVID-19 in Costa Rica Chaves et al. (2020) used a time varying reproduction number based on a serial interval to assess the impact and compared the study of Costa Rica with other countries Panama and Uruguay. A statistical framework to fit the syndromic and virological data was presented by Dorigatti et al. (2012), collected in Italy during the 2009-2010 A/H1N1 influenza pandemic using Susceptible- Exposed-Infected-Removed (SEIR) model with age structure. A stochastic SEIR transmission model with nine compartments based on symptomatic and asymptotic arising from zoonotic reservoirs and human to human trans- mission was modeled in Chowell et al. (2014). The study Kochańczyk et al. (2020) on COVID-19 used an SEIR model with multiple exposed states and analysed the dynamics of the epidemic based on the constant and time dependent contact rates. Bayesian approach with SIR model with correction of under-reporting was proposed by Oliveira et al. (2020).

The relative risk dengue fever occurrence was estimated by Samat and Percy (2012) while combining stochastic SIR model and Poisson distribution. In this study they assumed the infectives of the SIR model follow a Poisson distribution. The Markov Chain Monte Carlo method was used to find the posterior distribution. The number of infectives in the in-homogeneous population was predicted by Ristic et al. (2009) using Cramer-Rao lower bound (CRLB) method. The dynamics of COVID-19 based on Susceptible, Unquarantined infected, Quarantined infected and Confirmed (SUQC) infected was developed by Zhao and Chen (2020). Susceptible, Infected, Quarantined and Recovered (SIQR) model was used for analysing the dynamics of COVID1-9 in Brazil Crokidakis (2020) and in India Tiwari (2020). For estimating the occurrence of the diseases the epidemiologist frequently uses binomial, Poisson, or exponential distribution Flanders and Kleinbaum (1995). The authors Kucharski et al. (2020) estimated how the transmission rate of COVID-19 varied over time using sequential Monte Carlo simulation. Odhiambo et al. (2020) modeled the number of infectives of COVID-19 in Kenya as compound poisson distribution. The growth of COVID-19 pandemic in India is compared with several other countries in Ranjan (2020b).To predict the growth of COVID-19 in India the authors have used the SIR model and exponential model.

A Euclidean network with SIR model was used to predict the COVID-19 epidemic in China in Biswas et al. (2020). A time dependent SIR model was proposed by Chen et al. (2020). They have studied when the epidemic will end, how the asymptotic infections affect the spread and when the herd will be achieved. They have considered the detectable and undetectable infectives in the SIR model.

### *2.2. Flattening the Curve*

The spread of COVID-19 shows an exponential curve in the early stage of the pandemic and turns to power law function. The author in Verma et al. (2020) argued that the exponential function turns to power law function indicates the flattening of the infective curve. For the analysis nine countries were selected on the basis of a large number of positive cases. In the early stage of the pandemic the infective curves of all nine countries exhibited an exponential curve. And after the exponential curve they also exhibited polynomial function. The standard model of the SEIR model was modified and used in the study McBryde et al. (2020). Different reproduction number based on the different scenarios of Australia during the epidemic was used in this study to flatten and suppress the curve of COVID-19. Flattening of the curve using a logistic model is proposed in Ranjan (2020a); Ma (2020). Forming a logistic function by taking the maximum number of people that can be infected Allain (2020) illustrated the flattening of the COVID-19. The SEIR model was extended by including the on and off strategy of lock-down and neural regressor was also employed along with that to flatten the curve COVID-19 in Brazil Tarratacaa et al. (2020).

A mathematical model of geometric progression based on reproduction number and incubation period was modelled for the growth of COVID-19 Aluko (2020). Both static and dynamic reproduction numbers are employed in this model. Córdova-Lepe et al. (2020) used the geometric progression model to illustrate the cubic projection and exponential projection and suppression of COVID-19 in Chile.

Thus from the approaches that have been used in these articles we are using the SIR model for modelling the events during pandemic. We combine the SIR model with the exponential distribution and use Bayesian inference for modeling the mass gathering. For flight based event modeling we use the multiple linear regression and find the effect of these cases. For flattening the curve of COVID-19 we are using a geometric progression with multiple dynamic reproduction numbers based on fixed time intervals.

## 3. Epidemic Modeling in terms of Events

The world is facing a pandemic since the last year except Antarctica and all of them are in a race to slow down the spread of the virus. We focus on India as the cases are skyrocketing aiming to top the world. Due to the poor rate of illiteracy and economic conditions, people are avoiding the restrictions imposed by the government and hereby, increasing the number of cases in India. Mass gatherings are a part of Indian culture and gatherings around the different stages of pandemic violating the principle of social distancing aggravated the cases of the virus. During the initial stages of the pandemic large events with a turnaround of about 9000 people took place. Post gathering, there was a spike in the positive cases of coronavirus in the country. It is found that many positive cases were linked to that event. This is not the only cause of increase in the number of virus cases in India to rise. There are also other factors that affected the virus cases in India. The flights to and from India were banned by the government during the early stage of the pandemic as part of lock-down. There were many Indians in other countries who were affected by the pandemic. Thus Indian government came with a mission to bring back Indians from abroad. Thus the flights were resumed as part of this mission. As the number of flights and people came back to India the number of infectives also increased day by day. This paperconcentrated on modeling these two factors for the rise in the number of infectives.

### *3.1. Methodology*

#### *3.1.1. SIR Model*

There are many deterministic epidemiological models that can be used to model the spread of disease. In this study we are using a simple dynamic epidemic SIR model. The population is divided into three compartments: Susceptibles ($S$) is the number of susceptible individuals, Infectives ($I$) is the number of infected individuals, Removed infectives ($R$) is the number of recovered/dead individuals. The progression of one class to another class is shown in Fig. 1. The population moves from one compartment to another. Let $N$ be the population and the fraction of the total population in each of the three categories are

$$s(t) = S/N \tag{1}$$

$$i(t) = I/N \tag{2}$$

$$r(t) = R/N \tag{3}$$

(4)

where $t$ is the time in days. The differential equations of these classes is given by

$$dS/dt = -\beta S(t)I(t) \tag{5}$$

$$dI/dt = \beta S(t)I(t) - \gamma I(t) \tag{6}$$

$$dR/dt = \gamma I(t)) \tag{7}$$

Newly infected cases depend on the susceptible and already infected individuals. Thus susceptibles can be written in the form of Eq. (5) and the minus indicated that it is a decreasing variable. $\beta$ is the rate of infection and $\gamma$ is the removal rate and these two terms are varying functions. The removed variable depends on the Infectives and removal rate. Combination of Eq. (5) and 7 results in the infectives as in Eq. (6). The infection and removal rate should be monotonically increasing and decreasing functionsrespectively.

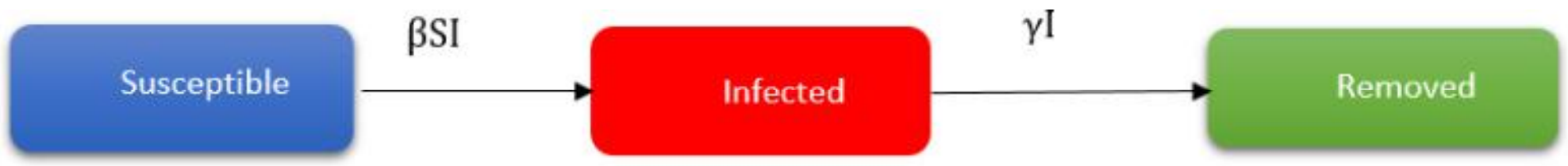

Figure 1: SIR model.

Since the population is very huge in India and infectives are more important than removed, we omit Eq. (5) and (7) from the system. That is we are considering only the infectives equation Zhong et al. (2020). Finite difference form of the infectives can be written as,

$$I(t + \delta t) = I(t) + (\beta - \gamma)I(t)\delta t \tag{8}$$

And the infectives and removal rate can be defined as,

$$\beta(t) = [I(t + \delta t) - I(t)]/I(t)\delta t \tag{9}$$
$$\gamma(t) = [R(t + \delta t) - R(t)]/I(t)\delta t \tag{10}$$

We are taking the cumulative number of infectives for the analysis. So beta rate and gamma rate can be calculated using infectives and removed on the infectives at $t$ time and $t + 1$ time respectively. Because beta rate is the rate of new infectives and gamma rate is the rate of removed individuals. By taking the difference of the infectives and removed population for a time $\delta$ and dividing it with the number of infectives gives the beta rate and gamma rate respectively.

#### *3.1.2. Event Modeling - I*

The number of confirmed cases due to COVID-19 is rising day by day and it exhibits an exponential curve. Therefore it is assumed that the number of infectives follows an exponential distribution. The probability distribution function of an exponential distribution is given by,

$$f(t) = \lambda\, e^{-\lambda t} \tag{11}$$

where $\lambda$ is the event rate. In this case for the infective rate $\lambda$ can be expressed by total number of infectives, number of days, and the total population. If $N$ is the total population, $I$ is the total number of infectives and $T$ is the number of days or observations then $\lambda$ can be defined as,

$$\lambda = \frac{I}{NT} \tag{12}$$

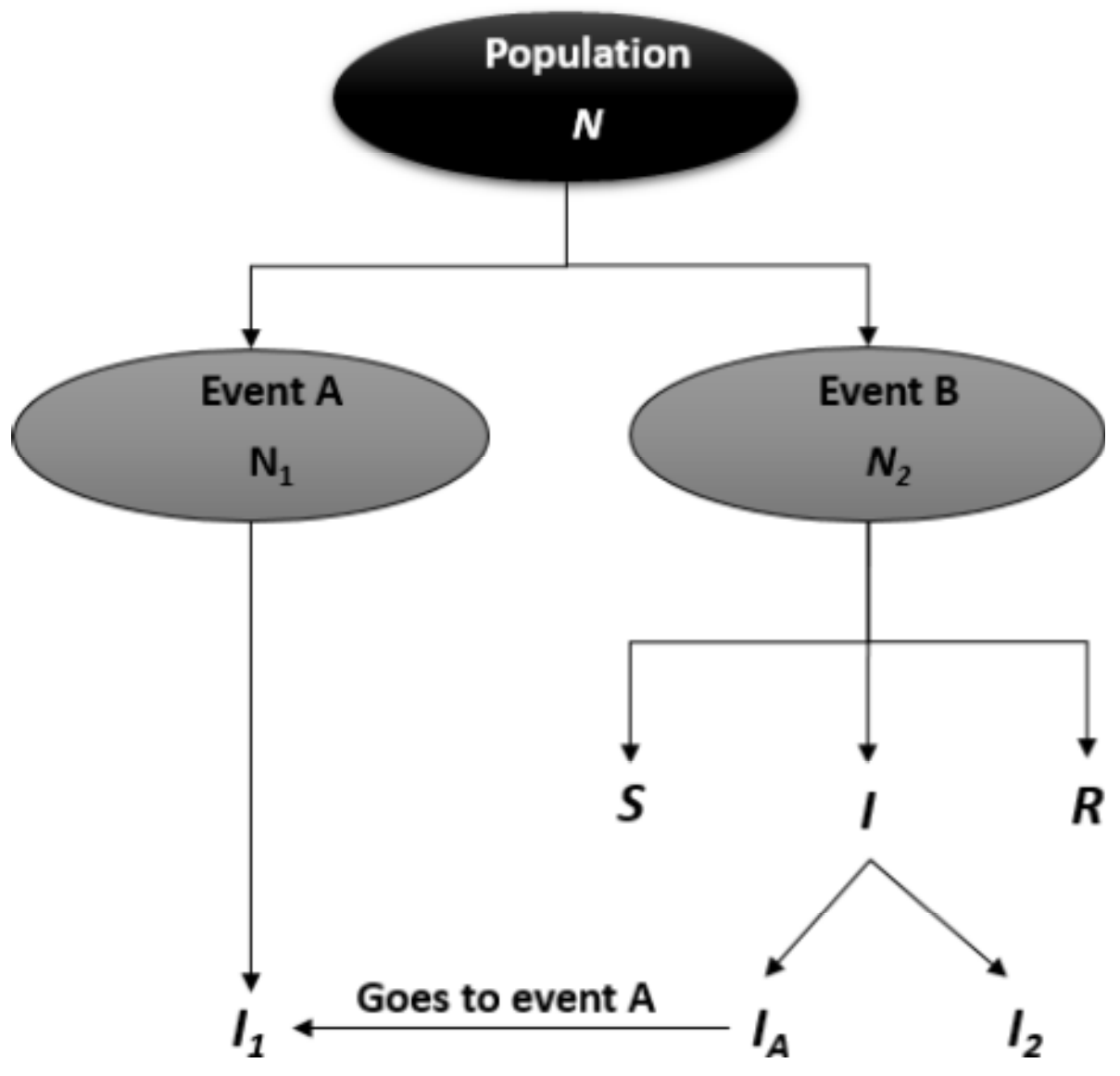

Figure 2: Proposed Events Model-I.

Fig. 2 shows the events in the proposed model. Let the population of the country be $N$ . Here we are considering two events. Event A is the people who attended the event having population $N_1$ and event B are the people who have not attended the event with population $N_2$. $I$ is the total infectives. The infectives $I$ can be divided into $I_A$ and $I_2$ where $I_A$ is the number of infectives that have attended the event or infectives that are linked to the event. Thus they move to event A and become $I_1$. And the remaining infectives $I_2$ is the infectives who are not a part of event A. Thus,

$$N = N_1 + N_2 \tag{13}$$

$$I = I_1 + I_2 \tag{14}$$

To find the effect of event A on B we use the Markov Chain Monte Carlo (MCMC) model. For estimating the posterior distribution of a parameter of interest by random sampling in a probabilistic space, MCMC is used van Ravenzwaaij et al. (2018). Formally, Bayes' rule is defined as

$$P(A/B) \propto P(B/A)P(A) \tag{15}$$

where $P(B/A)$ is the likelihood function, $P(A)$ is the prior probability and $P(A/B)$ is the posterior probability that has to be estimated. The Bayes' rule can be defined as,

$$P(A/B) = \frac{P(B/A)P(A)}{P(B)} \tag{16}$$

where $P(B)$ is the marginal distribution which is used to normalise the distribution. $P(A)$ and $P(A/B)$ belong to the same probability distribution family which is called a conjugate distribution. And the prior is called conjugate prior. Thus $P(B)$ can be defined as,

$$P(B) = \sum P(B/A)P(A) \tag{17}$$

Let $\lambda_1$ and $\lambda_2$ be the infective rate of events A and B respectively. Then,

$$P(A) = \lambda_1 e^{-\lambda_1 t} \tag{18}$$

$$P(B/A) = \lambda_2 e^{-\lambda_2 t} \tag{19}$$

$$P(B) = \Sigma \lambda_1 e^{-\lambda_1 t} \lambda_2 e^{-\lambda_2 t} \tag{20}$$

Thus,

$$P(A/B) = \frac{\lambda_1 e^{-\lambda_1 t} \lambda_2 e^{-\lambda_2 t}}{\Sigma \lambda_1 e^{-\lambda_1 t} \lambda_2 e^{-\lambda_2 t}} \tag{21}$$

Eq. (21) gives an estimation of the population affected by the event.

### 3.1.3. *Event Modeling - II*

Even though the number of infectives follows an exponential distribution, it also follows a multiple linear regression Uyanık and Güler (2013) also. Thus the number of infectives can be expressed as linear equation in terms of number of flights Odhiambo et al. (2020), people and days as follows Zaiontz (2014); Giuliani et al. (2020)

$$log\ I = a_1 + a_2x_1 + a_3x_2 + a_4x_3 \tag{22}$$

$$I(\mathbf{x}) = e^{a_1 + a_2x_1 + a_3x_2 + a_4x_3} \tag{23}$$

Where $x_1$ is the number of days or observation number, $x_2$ is the daily number of flights, $x_3$ is the number of people in those flights and $a_i$ are the parameters to be estimated. The above equation can be written in vector form as,

$$log\ I = XA \tag{24}$$

$$log \begin{bmatrix} l_1 \\ l_2 \\ . \\ . \\ l_3 \end{bmatrix} = \begin{bmatrix} 1 & x_{11} & x_{21} & x_{31} \\ 1 & x_{12} & x_{22} & x_{32} \\ . & . & . & \\ . & . & . & \\ 1 & x_{1n} & x_{2n} & x_{3n} \end{bmatrix} \begin{bmatrix} a_1 \\ a_2 \\ a_3 \\ a_4 \end{bmatrix} \tag{25}$$

where $X$ is the design matrix and vector A contains the regression coefficients. Let estimated values of fitted models be $\hat{I}(\mathbf{x})$. Then $\mathcal{E}$ is the error term which is the difference of estimated values of fitted models from the observed values. This paper tries to minimize the error $\mathcal{E}$ using Bayesian Optimization.

We denote the set of possible statistics as $\chi$ where $\chi \subseteq \mathbb{R}_4$ given x $\in \chi$ while solving for minimum norm solution.

$$\mathcal{E} = \min_{x \in \chi} \| I(x) \ - \ \hat{I}(\mathbf{x}) \|^2 \tag{26}$$

### 3.1.4. *Flattening The Curve*

The pandemic cases in India have been increasing rapidly and to flatten the curve we should know the progression of the growth of the virus. The growth of the coronavirus spread is similar to a geometric progression with a common ratio as shown in Fig. 3. The sum the geometric progression for common ratio greater than one can be written as,

$$S_n = a(r^n - 1)/(r - 1) \tag{27}$$

where $a$ is the initial term, $r$ is the common ratio and $n$ is the number of terms in the series.

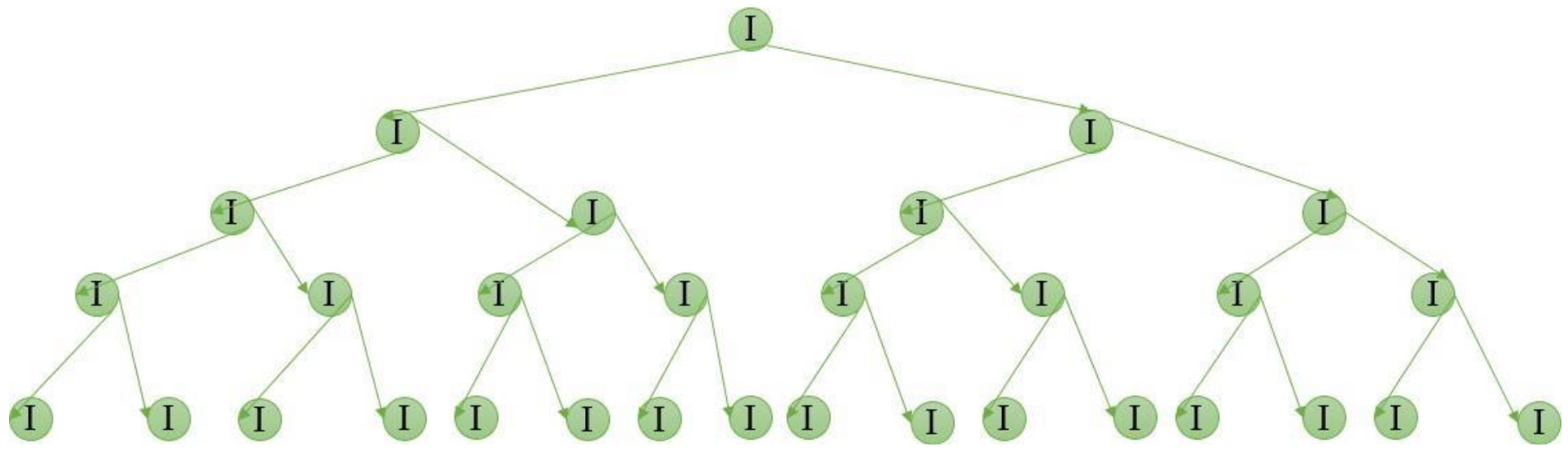

Figure 3: Geometric Progression of Infectives

With the assumption of a virus infected person arriving in a country, can be a cause of other infectives in the neighborhood. The next set of people can infect another set of people in a geometric progression series. The spread of these viruses depend upon the reproduction number $\rho$. A reproduction number is the average number of infectives that can be produced by a single infective and these constitute the average number of secondary cases Delamater et al. (2019). In the proposed scenario case the reproduction number becomes the common ratio.

When these infectives are exposed to other people, there is a delay in showing the symptoms of infection in the latter case and that time delay is known as incubation period Patrikar et al. (2020). It can be expressed as $\tau$. We have taken the initial term $a$ as 1 and $t$ the number of days as $n$. Thus the cumulative sum of infectives can expressed as,

$$S_t = \frac{\rho^{t/\tau} - 1}{\rho^{1/\tau} - 1} \tag{28}$$

The reproduction number changes over time as we are taking precautions to slow down the growth of COVID-19 cases. When the reproduction number becomes less than one, then only the curve of COVID-19 flattens. Thus here we are taking dynamic reproduction numbers for each fixed time interval. Let $\rho_1$ be the reproduction number for first $t+1$ days, $\rho_2$, be the reproduction number for the next $t+1$ days, $\rho_3$ be the reproduction number for the next $t+1$ days, and $\rho_k$ be the reproduction number for the last $t+1$ days. Thus now we can write the cumulative sum of infectives with dynamic multiple reproduction number based on fixed time interval as,

$$S_t = \frac{\rho_1^{t/\tau} - 1}{\rho_1^{1/\tau} - 1} + \rho_1^{t/\tau} \rho_2^{1/\tau} + \frac{\rho_2^{t/\tau} - 1}{\rho_2^{1/\tau} - 1}$$
$$\rho_1^{t/\tau} \rho_2^{t+1/\tau} \rho_3^{1/\tau} \frac{\rho_3^{t/\tau} - 1}{\rho_3^{1/\tau} - 1} + \ldots\ldots + \tag{29}$$
$$\rho_1^{t/\tau} \rho_2^{t+1/\tau} \rho_3^{t+1/\tau} \ldots\ldots \rho_k 1/\tau \frac{\rho_k^{t/\tau} - 1}{\rho_k^{1/\tau} - 1}$$

## 4. Results

The data of COVID-19 in India was collected from kaggle SRK et al. (2020) and the data for flights was collected from the Ministry of External Affairs Ministry of External Affairs Government of India (2020). From the data collected we have shown the infected and removed curves as shown in Fig. 4. More than 5.9 million people are infected in India and more than three quarters of them are removed. With that removed, only 94503 people are dead till $26^{th}$ September, 2020 and the remaining 4.9 million are cured. From the experiments conducted only three percent of the total infectives in India have lost their life and about 83 percent of the infectives have been recovered from the infection.

The state-wise confirmed cases of COVID-19 are plotted in Fig. 5. The most affected state of COVID-19 is Maharashtra with about 1.2 million confirmed cases. And out of that more than 1 million people have been recovered and about 35000 people are dead. After Maharashtra there comes Andhra Pradesh with more than 646000 infectives. The least affected state is Mizoram with 1835 confirmed cases.

Fig. 6 shows the infectives and removed curves of each state and union territory. And it clearly says that no state has escaped from the COVID-19 infectives. They are recovering but the infectives in these states exist.

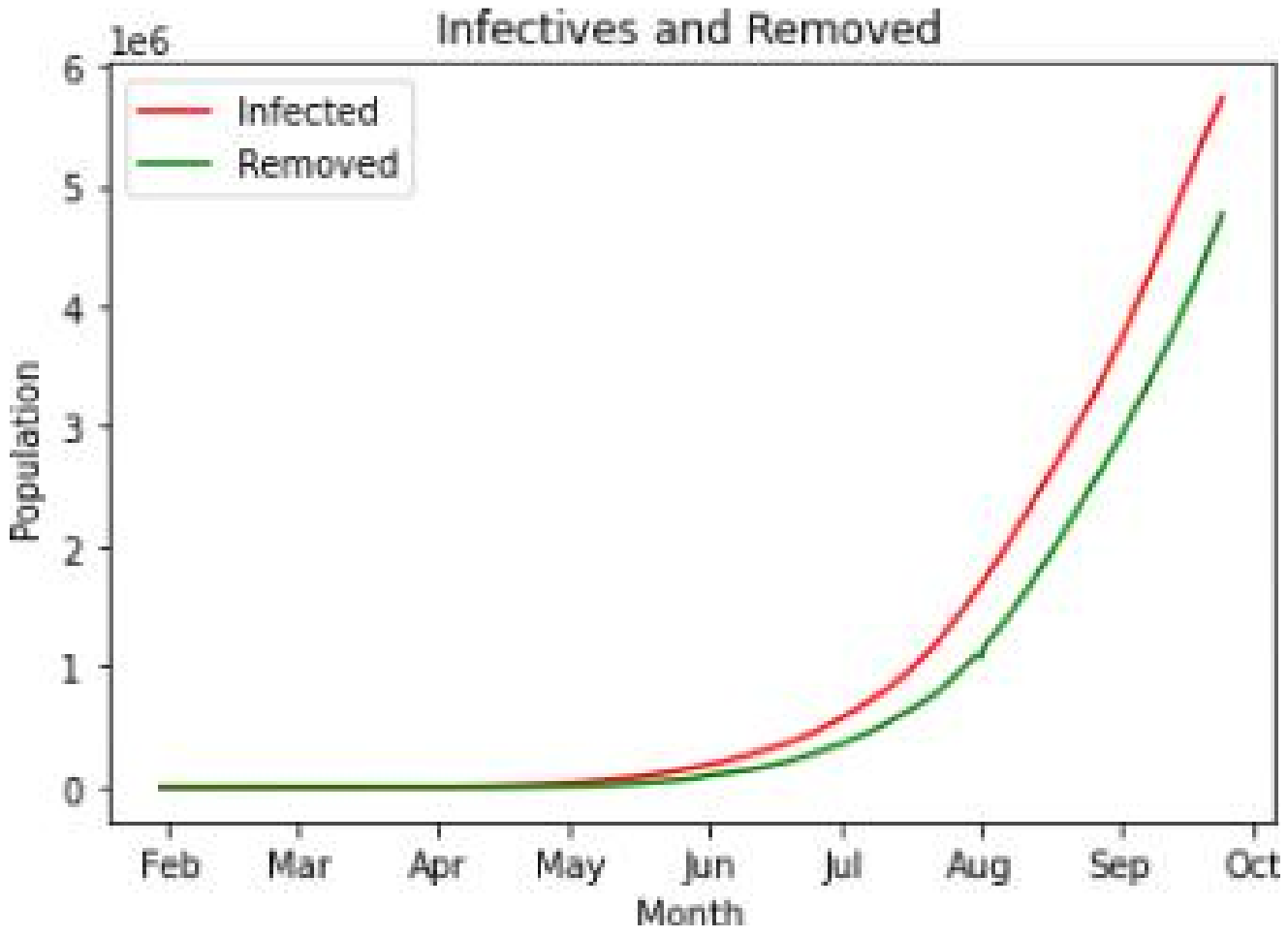

Figure 4: Infectives and Removed

The beta rate of the infectives should be a monotonically decreasing function. Even though the number of infectives is increasing daily, the infection rate has decreased. That is, the beta rate has decreased. Fig. 7 shows the beta rate of India. The blue dots show the actual beta rate and the red line shows beta rate fitted using linear regression. And the fitted beta shows a monotonically decreasing function. Till now beta rate has not reached zero. But it clearly shows that eventually beta rate will reach zero and there will be no more new infectives in India.

Fig. 8 shows the infectives of the event A. In the early stage of pandemic in India there were only few infectives in India. After the event, there was a rapid increase in the number of infectives. It was found that there were 4291 infectives linked to the event till $18^{th}$ April, 2020. Fig. 8 shows that there are 8932 people who got infected in the event till then. It clearly indicates that the event has positively affected the spread of COVID-19 in India.

On $7^{th}$ May, 2020 the flights were resumed to bring back Indians from abroad. By using generalized linear regression, the multiple linear equation of Eq. (3.21) in terms of number of days, flights and people arrived can be written as,

$$log\ I = 10.8478 + -0.0027x_1 + 0.00002007x_2 + 0.0499x_3 \quad (30)$$

By taking the exponential of Eq. (30), we can get the number of infectives. Thus the multiple linear model can be fitted as shown in Fig. 9. From this

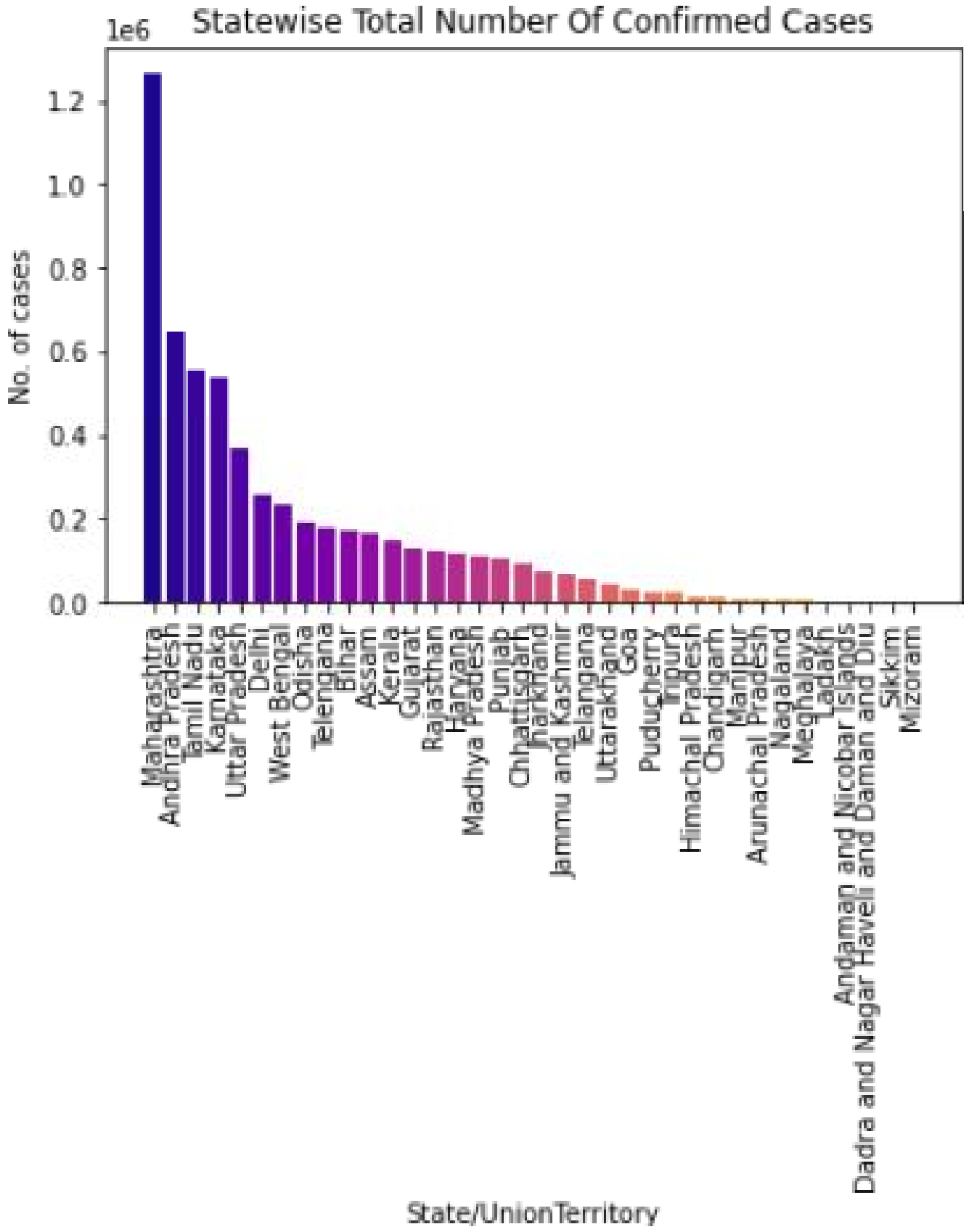

Figure 5: State-wise Infected

figure, we can understand that the fitted data and the actual data moves parallel. The fitted data is almost equal to the actual data. Thus from the figure we can say that $x_1$, $x_2$, and $x_3$ are related to the number of infectives.

From the data, the summary of multiple linear equations based on the number of flights, passengers, and days using generalized linear regression is shown in Fig. 10. Since the null hypothesis values of the variables are less than 0.05 means that the number of days, number of flights, and people who came to India has a positive relation with the number of infections in India.

The geometric progression of COVID-19 is based on the reproduction number. Thus to calculate the reproduction number COVID-19 we are using the daily number infectives as in Aluko (2020). The first step is to calculate $x_i = c_i/c_{i-1}$ where $c_i$ and $c_{i-1}$ are number people infected on day $i$ and

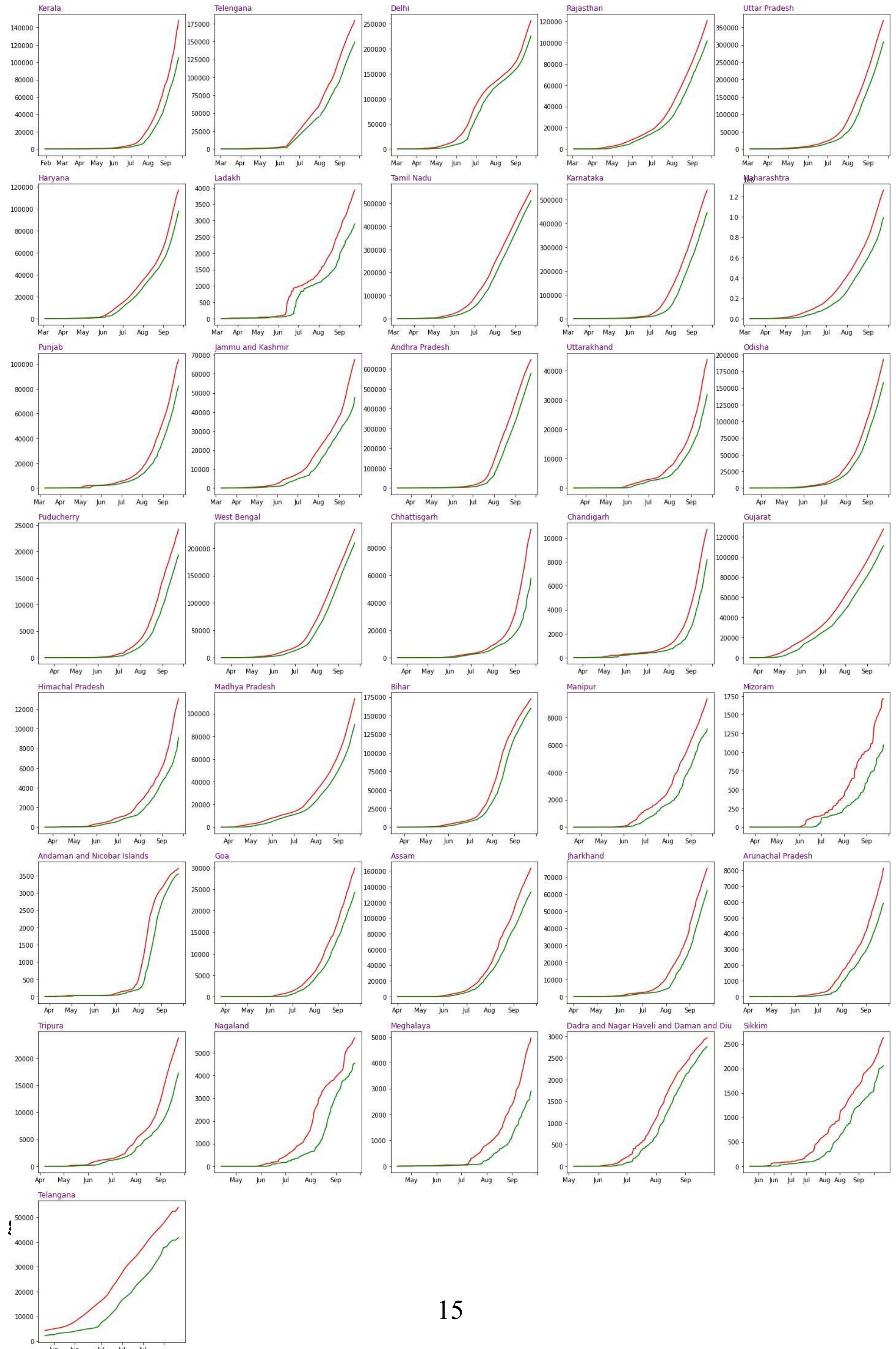
Kerala
Telengana
Delhi
Rajasthan
Uttar Pradesh
Haryana
Ladakh
Tamil Nadu
Karnataka
Maharashtra
Punjab
Jammu and Kashmir
Andhra Pradesh
Uttarakhand
Odisha
Puducherry
West Bengal
Chhattisgarh
Chandigarh
Gujarat
Himachal Pradesh
Madhya Pradesh
Bihar
Manipur
Mizoram
Andaman and Nicobar Islands
Goa
Assam
Jharkhand
Arunachal Pradesh
Tripura
Nagaland
Meghalaya
Dadra and Nagar Haveli and Daman and Diu
Sikkim
Telangana

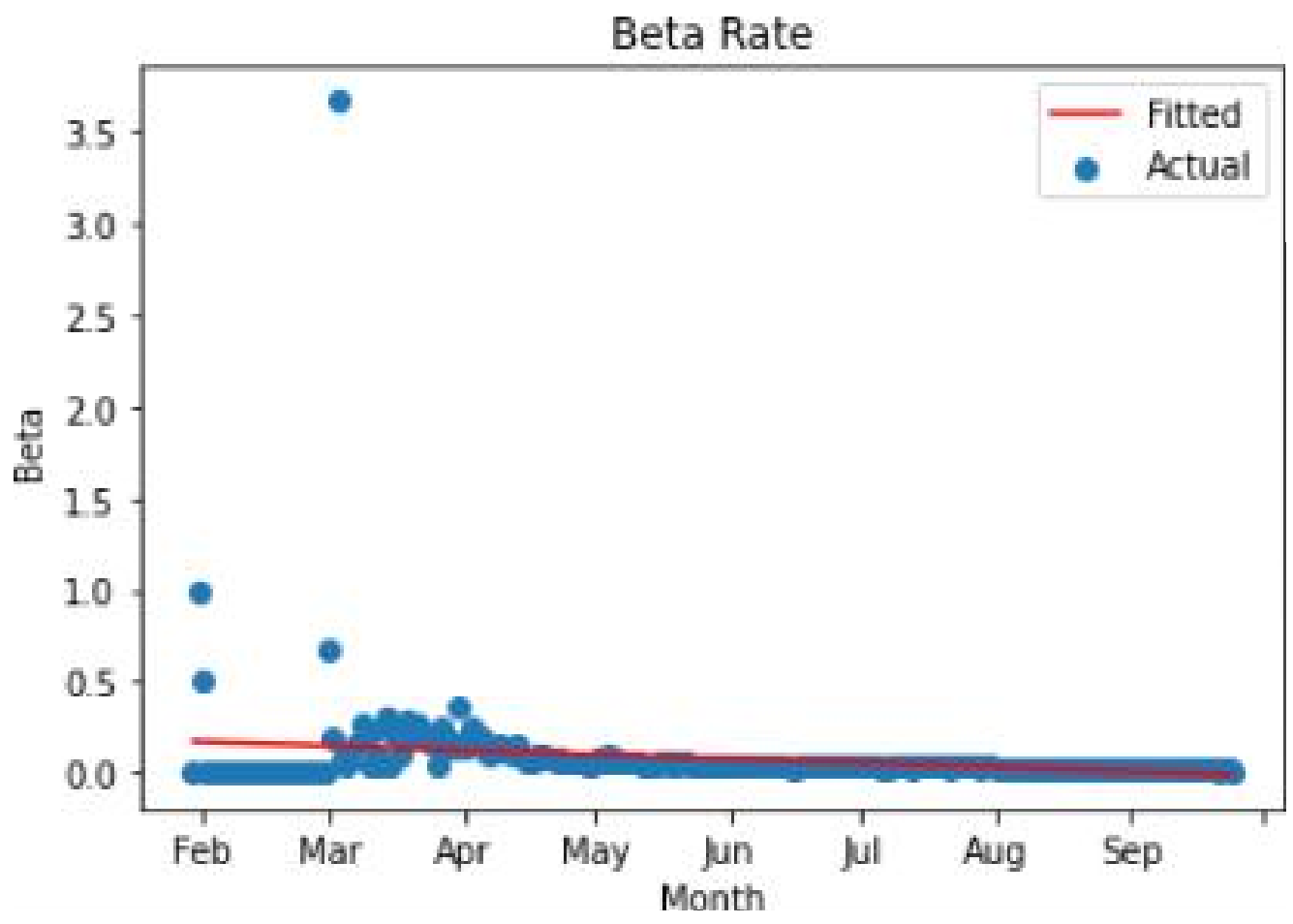

Figure 7: Beta Rate.

$i-1$ and $x^{\tau}$ gives the daily reproduction number. The maximum threshold for reproduction number is assumed as 3.5. The average of the reproduction number of each time period is taken as the reproduction number for that time interval. The average incubation period of COVID-19 is taken as 5.2 days. Here we are analysing growth for four different time intervals $t = 2, 7,$ 14 and 20 days of time gap.

In the time intervals that we have chosen, as the time interval increases the predicted value becomes more closer to the actual number of infectives. The semilogy graphs of infectives seem to be flattening. Even though it is increasing, the increasing rate of infectives has reduced. Since the reproduction rate has not reached below 1 new infectives will occur and when the reproduction number reduces to less than 1 it would flatten. In the Fig. 11 the predicted and actual number infectives are much different in the normal curve. But when it comes to semilogy curves, the difference seems less com- pared to the normal curves. And it depicts a kind of starting of flattening phase. Comparing the semilogy curves, the curve with time interval $t = 2$ is overestimating the number of infectives. But $t = 14$ and $t = 20$ is more close to the actual number of infectives. Since it shows a starting phase flattening it will take some more months to flatten the curve of the COVID-19 infectives.

In the study Zhong et al. (2020) using the subset data they are fitting the beta rate ($\beta$) linearly and exponentially. Since the experiment was conducted

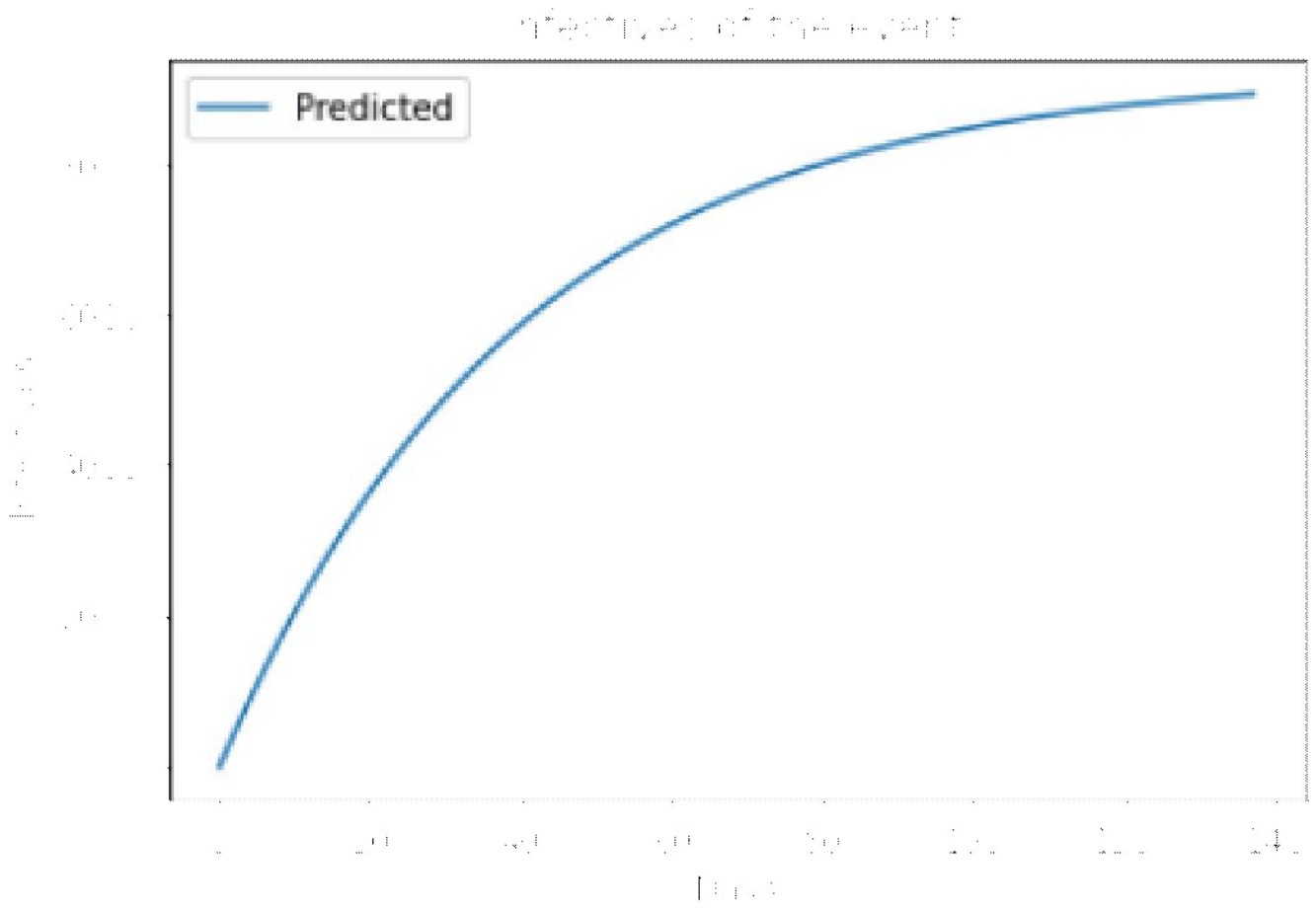

Figure 8: Infectives of Event A

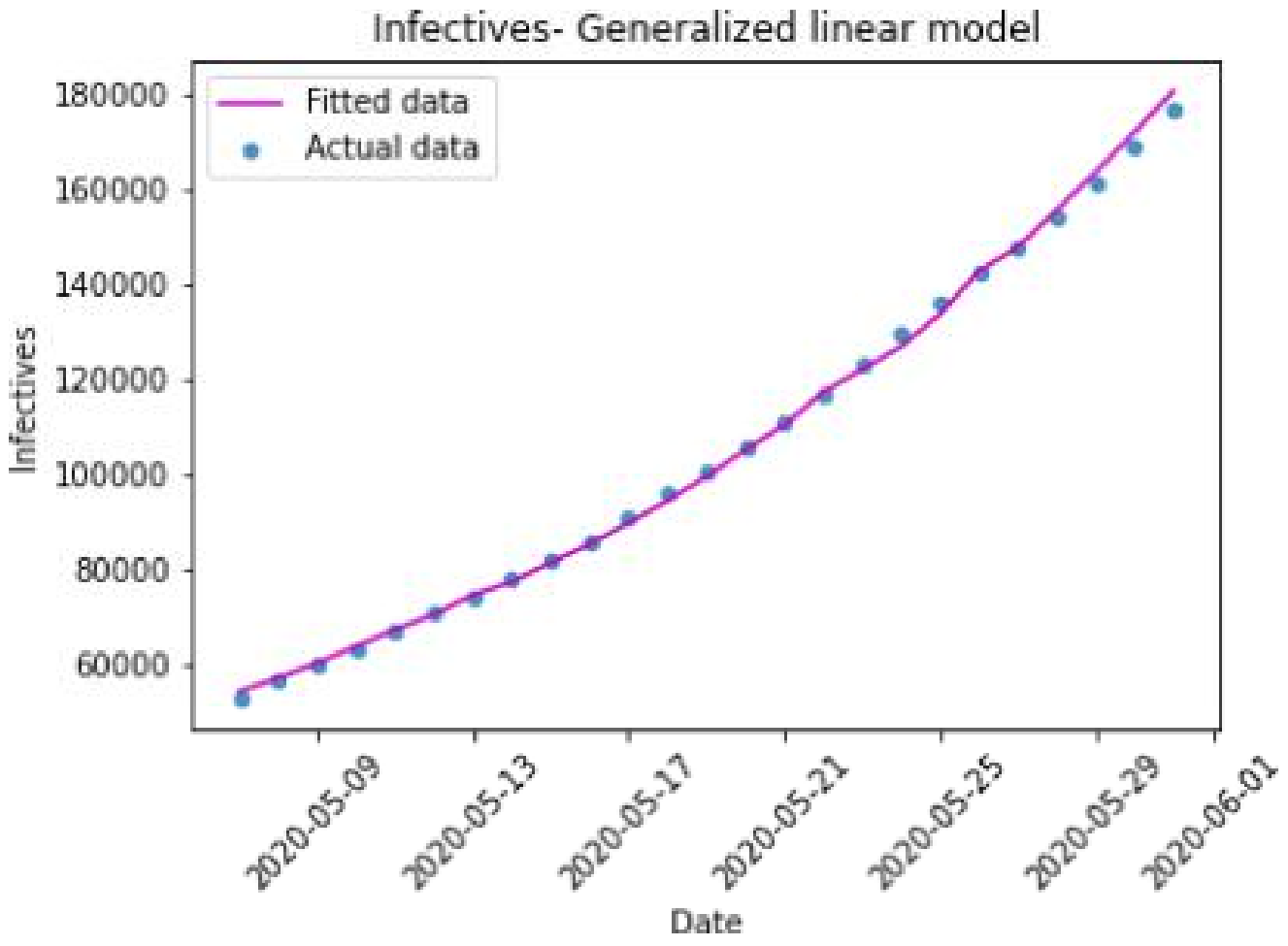

Figure 9: Generalized Linear Regression

Generalized Linear Model Regression Results

| | | | |
|---|---|---|---|
| Dep. Variable: | Infected | No. Observations: | 25 |
| Model: | GLM | Df Residuals: | 21 |
| Model Family: | Gamma | Df Model: | 3 |
| Link Function: | log | Scale: | 0.00017531 |
| Method: | IRLS | Log-Likelihood: | -213.07 |
| Date: | Wed, 17 Jun 2020 | Deviance: | 0.0036886 |
| Time: | 15:14:59 | Pearson chi2: | 0.00368 |
| No. Iterations: | 7 | | |
| Covariance Type: | nonrobust | | |

| | coef | std err | z | P>\|z\| | [0.025 | 0.975] |
|---|---|---|---|---|---|---|
| const | 10.8478 | 0.007 | 1510.161 | 0.000 | 10.834 | 10.862 |
| x1 | 0.0499 | 0.000 | 133.658 | 0.000 | 0.049 | 0.051 |
| x2 | -0.0027 | 0.004 | -0.708 | 0.479 | -0.010 | 0.005 |
| x3 | 2.007e-05 | 1.89e-05 | 1.062 | 0.288 | -1.7e-05 | 5.71e-05 |

Figure 10: Generalized Linear Regression Summary

in the early stage of the epidemic there was only a limited amount of data and it was easy to take the subset data and select the best from them according to the criteria for the beta rate. But in the proposed case we have a large amount of data, so it is difficult to take the subset of the data and select the best fit from the curves. Thus, here we have used only a single linear regression to fit the beta rate and gamma rate of the whole data. The fitted data is applied on the Eq. (8) and the curves we got are Fig. 12 and Fig.
13. From these figures it is also clear that the curve of infectives has not flattened. It is in the starting the phase of flattening.

This event based modeling can be used in other studies also. It is appli- cable to other pandemics or in other countries also. The events can be based on ages, traffic, area, and many other cases.

## 5. Conclusion

Even though the pandemic has affected the whole world, it has affected India drastically. India has come to second position in the world with most number of COVID-19 cases reported. Due to economic slow down the restrictions have reduced and it paved a way to increase the cases in India. When compared to other countries a low fatality rate and recovered cases are recorded. In this study we have proposed an event based model using exponential distribution and Bayesian approach for the mass gathering event. A generalised linear regression model is used to model the number of flights arrived and the number of infectives. With the study of the progression series the number of infectives is still in an increasing mode and it has not reached the flattening stage.

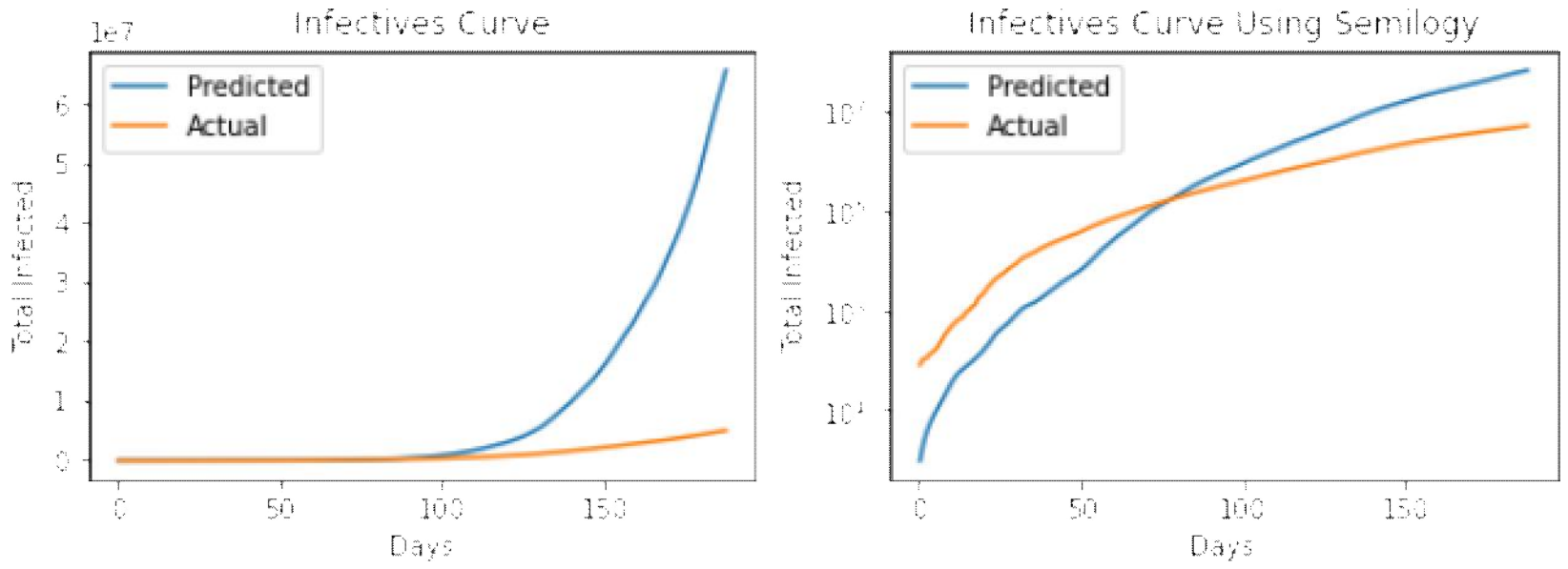

(a) When t=2

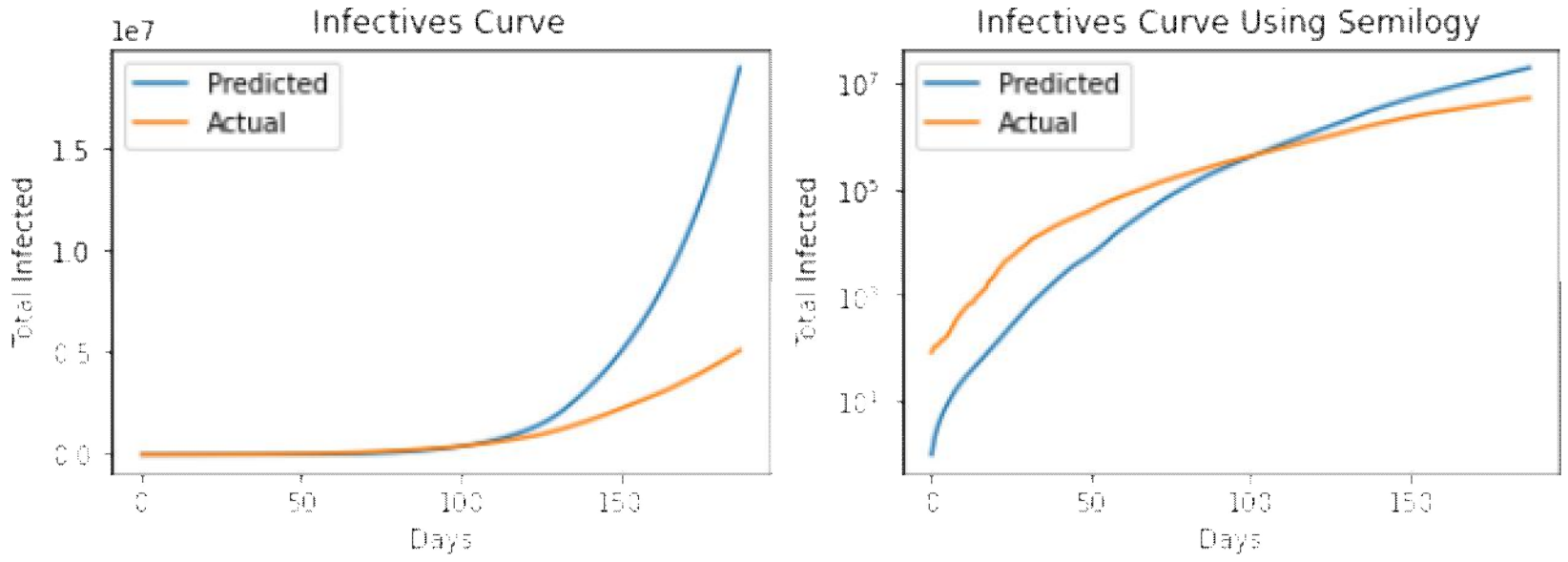

(b) When t=7

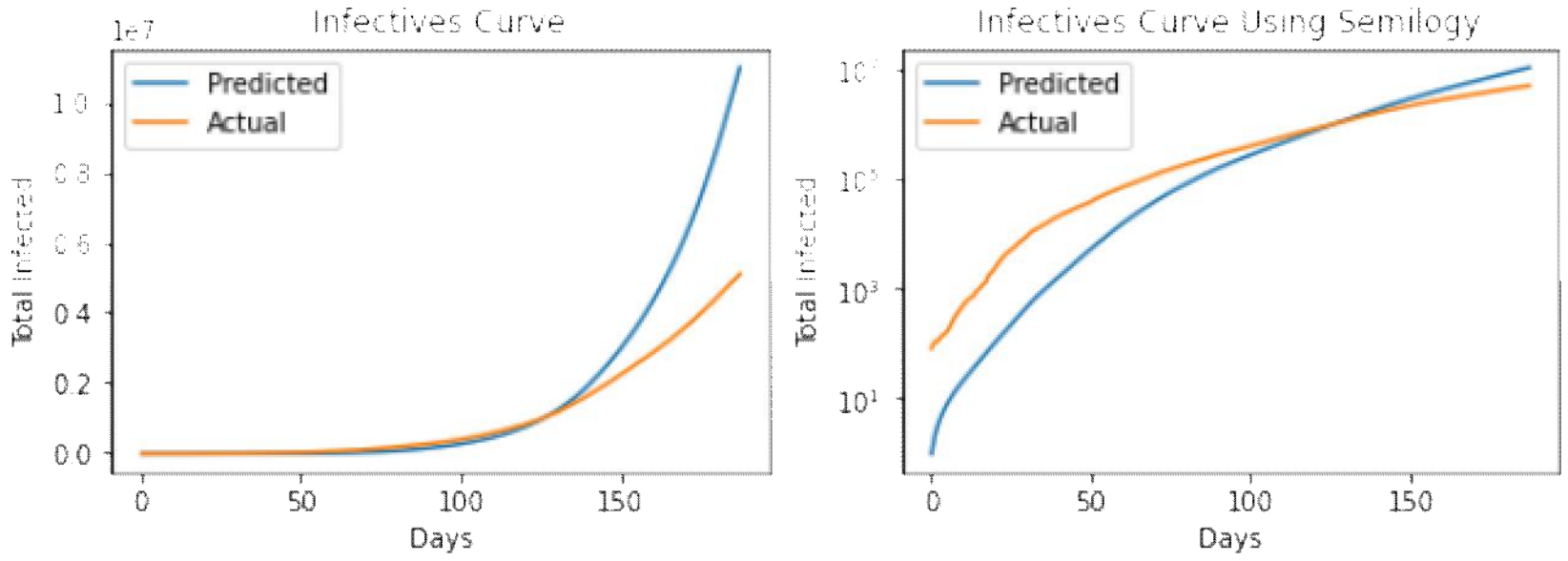

(c) When t=14

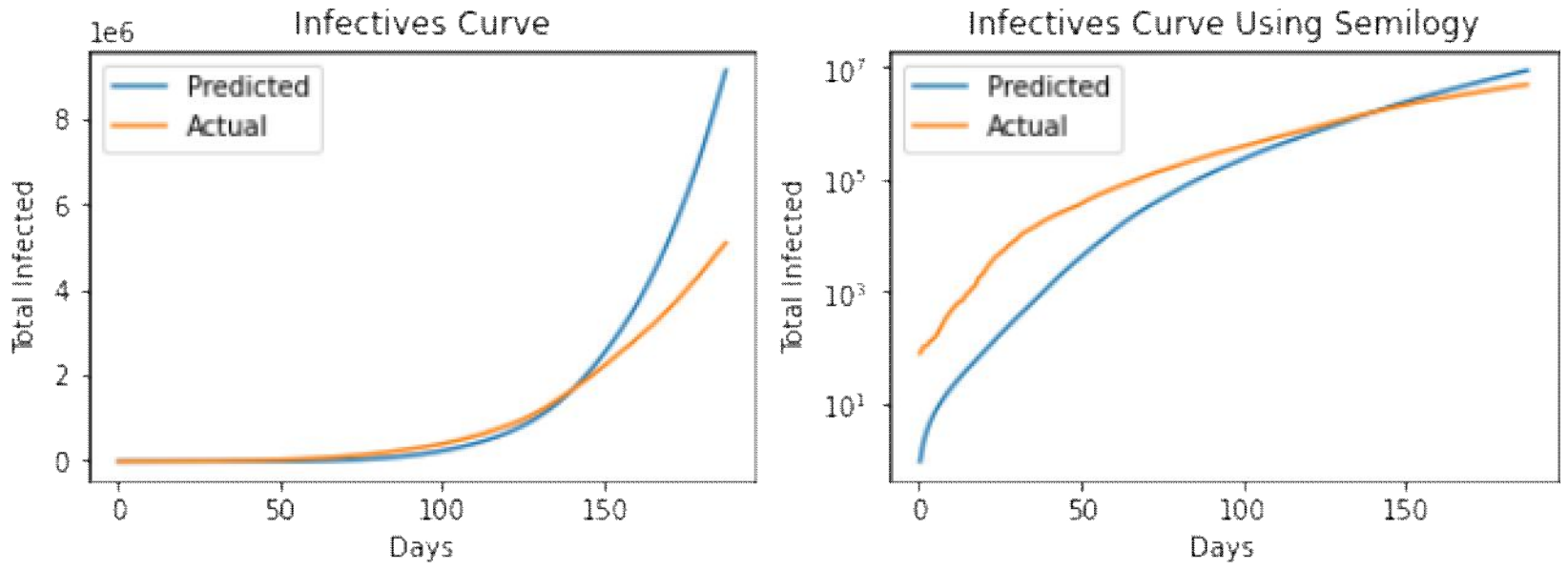

(d) When t=20

Figure 11: Geometric Progression of COVID-19 based on Fixed Time Interval Reproduction Number

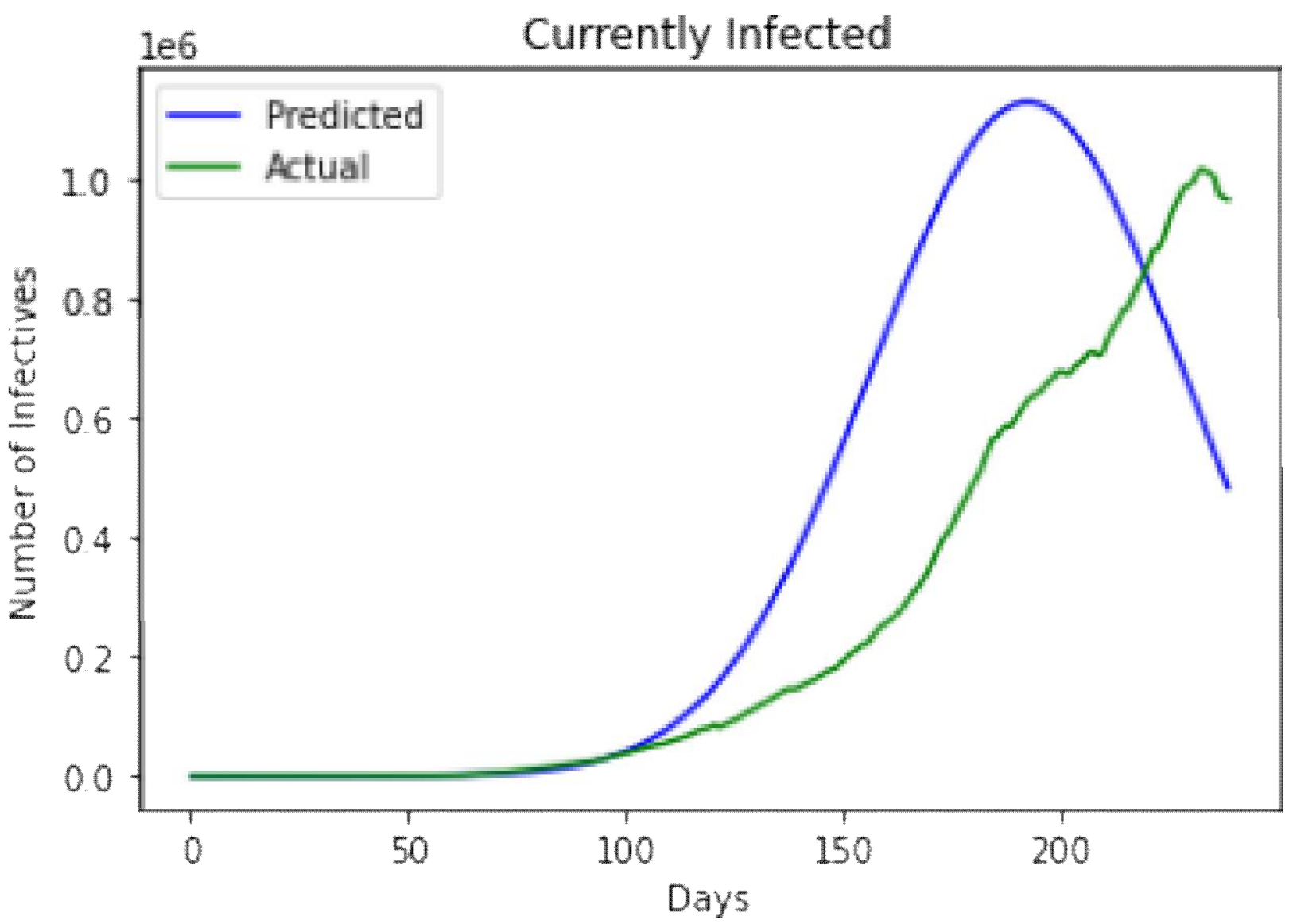

Figure 12: Infectives Curves based Beta and Gamma Rate

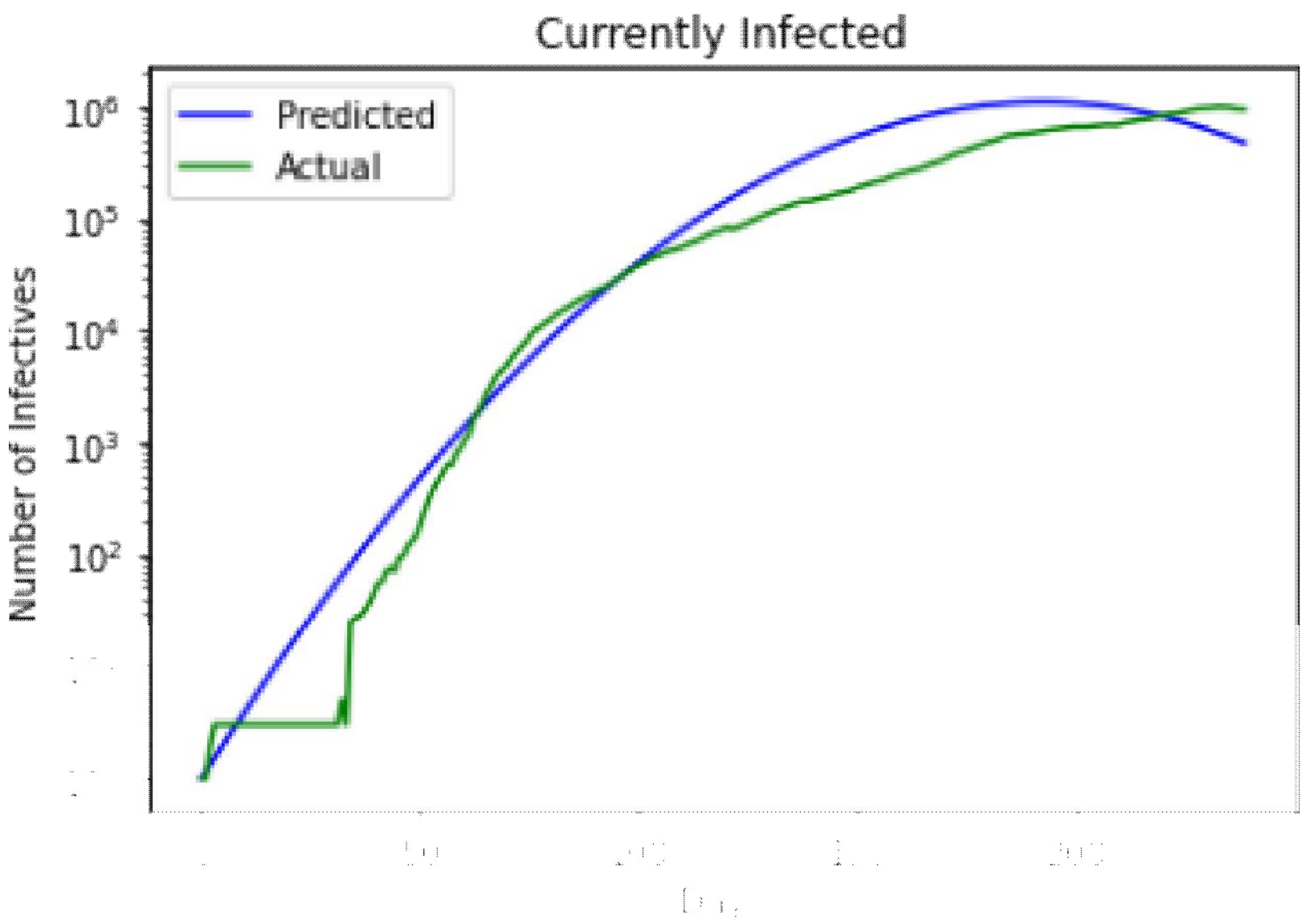

Figure 13: Infectives Semilogy Curves based Beta and Gamma Rate